%% file: main.tex
\documentclass[a4paper]{sig-alternate}
\usepackage[english]{babel}
\usepackage[utf8x]{inputenc}
\usepackage[T1]{fontenc}
\usepackage{setspace}
\usepackage{amssymb}  % assumes amsmath package installed
\usepackage[linesnumbered,lined,ruled, vlined, boxruled]{algorithm2e}
\usepackage{mathtools}
\usepackage{thmtools, thm-restate}
\usepackage[noend]{algpseudocode}
\usepackage{cite}
\usepackage{soul}

\usepackage{amsmath}
\usepackage{graphicx}
\usepackage{subfigure}
\usepackage{caption}
\usepackage[colorlinks=true, allcolors=blue]{hyperref}

\title{Detection of AQM on Paths using Machine Learning Methods}

\numberofauthors{3}
\author{
\alignauthor
Cenk Baykal
       \email{baykal@mit.edu}
\alignauthor
Wilko Schwarting 
       \email{wilkos@mit.edu}
\alignauthor
Alex Wallar 
       \email{wallar@mit.edu}%
}

\begin{document}
\maketitle

% Problem Definition.
\input{problem}

% Approach.
\input{approach}

% Evaluation.
\input{evaluation}

% Conclusion.
\input{conclusion}

\bibliographystyle{unsrt}
\bibliography{sample}

\end{document}

%% file: problem.tex
\section{Problem Definition}
\label{sec:problem}
Active Queue Management (AQM) schemes, such as PIE \cite{pie} and RED \cite{red}, have been particularly effective in combating bufferbloat and attenuating network congestion. Despite their widespread success and improved performance over traditional Tail Drop queuing schemes, 
the extent of AQM deployment in contemporary networks is not known \cite{tada}. Detecting the presence of AQMs in a network has potential to not only provide insight into a network's internal characteristics (i.e., network tomography), but also facilitate network optimization \footnote{e.g., in the case that a transport protocol is known to perform poorly over an AQM or certain subset of AQMs \cite{tada}.}.

In this paper, we address the problem of determining whether a bottleneck router on a given network path is using an AQM or a drop-tail scheme. We assume that we are given a source-to-sink path of interest -along which a bottleneck router exists- and data regarding the Round-Trip Times (RTT) and Congestion Window (CWND) sizes with respect to this flow. We develop a reliable classification algorithm that solely uses RTT and CWND information pertaining to a single flow to classify the queuing scheme, Tail Drop or AQM, used by the bottleneck router. We evaluate our method and present results that demonstrate our algorithm's highly accurate classification ability across a wide array of complex network topologies and configurations.

%% file: approach.tex
\section{Approach}
\label{sec:approach}
Determining whether the bottleneck router in a network path is employing a drop-tail or AQM queuing scheme is fundamentally a classification problem, which has been rigorously studied in Machine Learning literature. To this end, we employ a data-driven approach
that entails generating random network configurations of varying complexity, simulating the generated scenarios using Mininet \cite{mininet}, and training a classification algorithm on the obtained data. Our approach contains three main components: (i) randomized network topology generation and simulation, (ii) feature engineering to encode relevant information, and (iii) training of the classification algorithm. 

\begin{figure}[!htb]
  \centering
  \includegraphics[width=\linewidth]{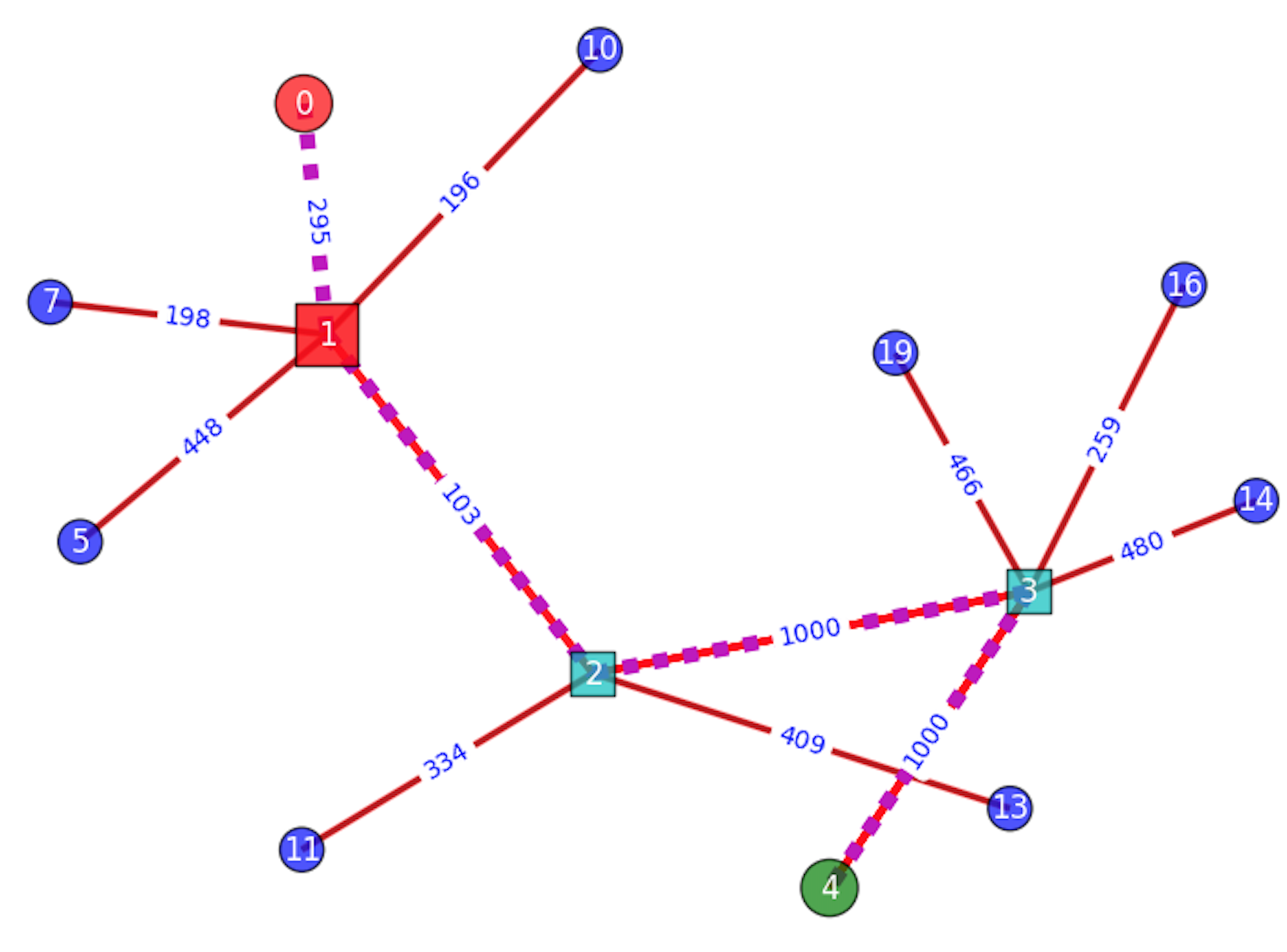}
  \centering
   \includegraphics[width=\linewidth]{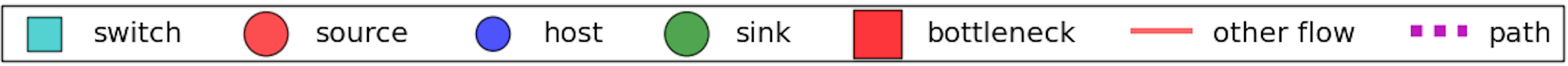}
\caption{A random network topology generated by our framework. Edge weights denote link capacities.}
\label{fig:topology}
\end{figure}

\subsection{Random Topology Generation}
Properly training a machine learning algorithm to simultaneously achieve high prediction accuracy and generalization to different kinds of network configurations requires a large and diverse set of training data. Hence, we opted to leverage randomization in order to generate a wide array of sufficiently diverse network topologies, which we subsequently simulated using Mininet. In particular, we developed a framework for constructing varied network configurations. 

An example of a random network topology $G = (V, E)$ generated by our framework is depicted in Fig.~\ref{fig:topology} and is constructed as follows. We initialize a \textit{path graph}, i.e., a graph defined by a path from the source node $s$ to the sink node $g$, $\pi = (s, v_1, \ldots, v_n, g)$, consisting of $n \sim \mathcal{U}(3,5)$ \footnote{$\mathrm{U}(a,b)$ denotes a continuous uniform distribution, whereas $\mathcal{U}(a,b)$ denotes a discrete uniform distribution.} switches, i.e., routers. For each switch along the path $v \in \{v_1,\ldots,v_n\}$, a random Erd\"{o}s-R\'{e}nyi graph $G_\text{rand}(n_p, p)$ \cite{erdos} characterized by the uniformly drawn number of nodes $n_p \sim \mathcal{U}(1,5)$ and connectivity probability $p \sim \mathrm{U}(0,1)$ is generated such that each node is a switch or a host with equal probability; the largest connected component $G_\text{conn} \subseteq G_\text{rand}$ is identified and connected to switch $v$ by an edge (i.e., network link). Additional $h \sim \mathcal{U}(1,5)$ hosts are generated and connected to vertex $v$.

Randomly chosen values for link delays and capacities are generated to diversify the network configuration. For each link $e \in E$, we set a random link delay $d \in \mathcal{U}(10,100)$ and link capacity $c \in \mathcal{U}(10,1000)$. To increase network complexity and simulate scenarios with bottlenecks caused by competing flows, we add auxiliary flows, i.e., paths, from randomly chosen host nodes to the sink node. The randomly generated link capacities are adjusted accordingly so that there exists a single bottleneck router along the path $\pi$ from source to sink, which is subsequently labeled as either Tail Drop or AQM (denoting the queuing scheme to be simulated).\footnote{The ranges used for the uniform distributions can be easily altered to generate random networks of desired complexity.}

\subsection{Feature Engineering}

\begin{figure}[t!]
\centering
\subfigure[Tail Drop]{%
  \includegraphics[width=\linewidth]{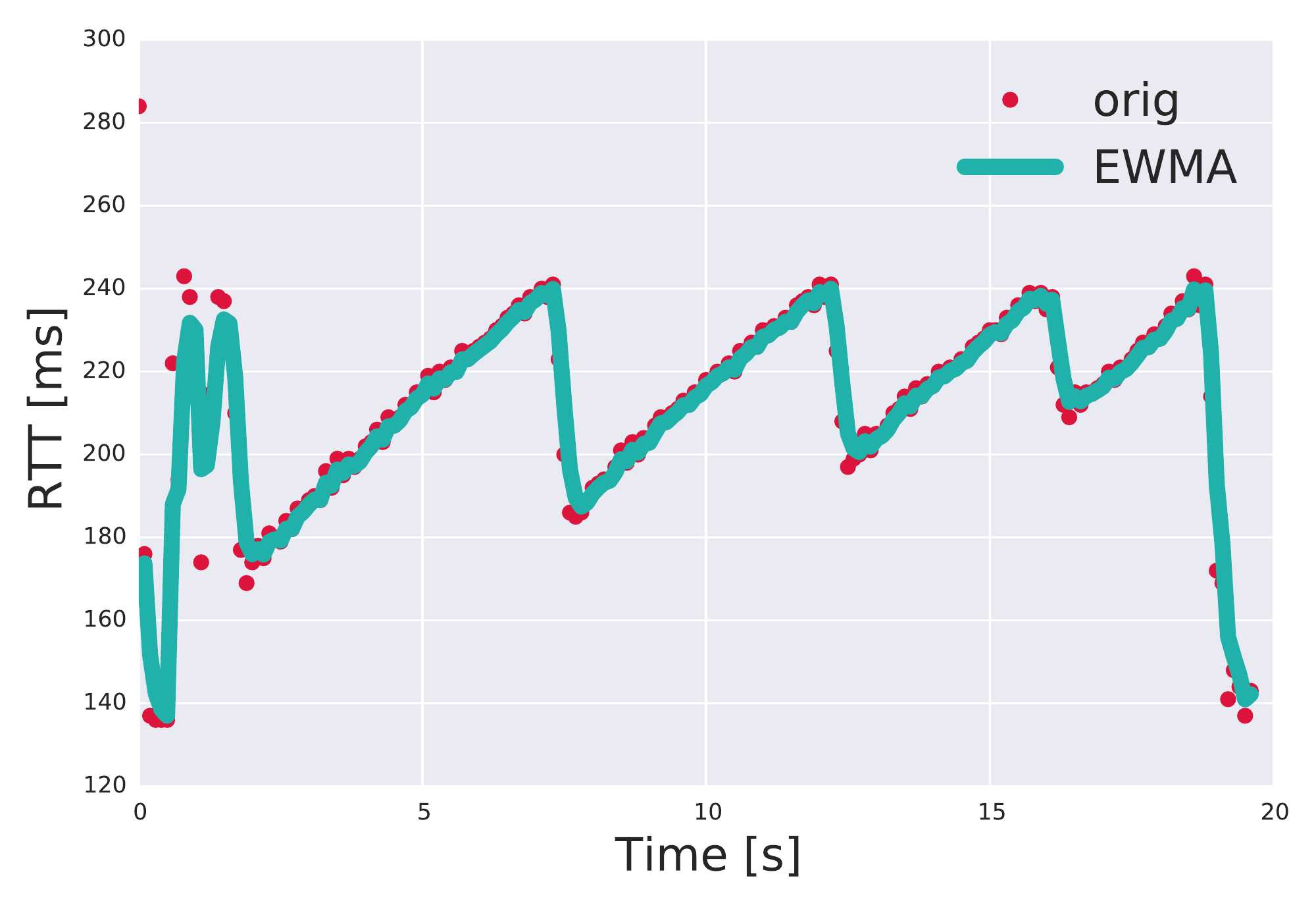}%
\label{fig:tail-drop}}
\subfigure[PIE]{%
   \includegraphics[width=\linewidth]{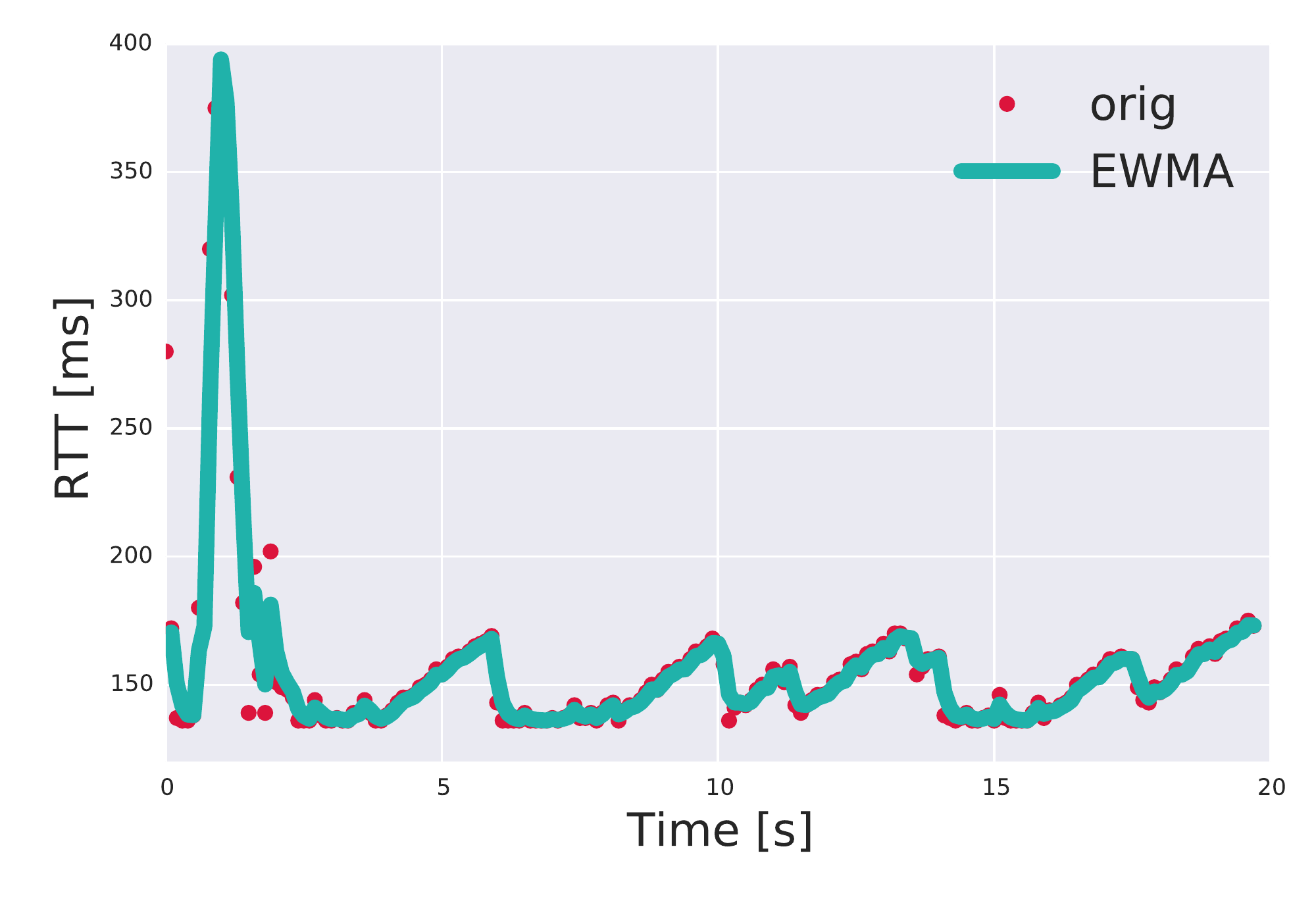}%
\label{fig:pie}}
\caption{RTT data obtained from two distinct Mininet simulations of the same network topology with the exception of the queuing scheme used by the bottleneck router: (a) Tail Drop and (b) PIE (AQM).}
\label{fig:data}
\end{figure}
The randomly generated network topologies are simulated using Mininet to obtain training data in the form of RTT and CWND information over time, as further detailed in Sect.~\ref{subsect:training}. Figs.~\ref{fig:tail-drop} and \ref{fig:pie} depict the RTT data gathered from a 20 second Mininet simulation of the topology shown in Fig.~\ref{fig:topology}. In order to construct a vector of distinguishing features, we examined a multitude of diversely-ranged RTT and CWND figures, similar to the figures shown above. Our observations indicated that high amount of noise was present in the data, therefore, we used the Exponentially Weighted Moving Average (EWMA) filter with parameter $\alpha=0.3$ to smoothen the resulting RTT and CWND data.

Our observations reaffirmed the inherent differences in the methodology of Tail Drop and AQM schemes: Tail Drop does not drop packets until the queue is full, which leads to a pronounced sawtooth pattern in the RTT graph (Fig.~\ref{fig:tail-drop}), i.e., lengthy periods of linear increase in the RTT plot followed by a sharp drop. PIE, on the other hand, preemptively (and probabilistically) drops packets, which results in relatively short bursts of increasing and decreasing RTT values. 

We combined our insights from experimental observations and the inherent differences in Tail Drop and AQM schemes to construct a vector of distinguishing features. Our feature vector was 72 dimensional (36 features each for RTT and CWND) that were computed for each simulated time series. Examples of particularly distinguishing features that constituted our feature vector include: statistics regarding the first and second order gradient (e.g., mean, variance, maximum), the number of local minima \& maxima, the $L_2$-norm squared total variation, the maximum sum subsequence, and the mean and variances of the local maxima. Fig.~\ref{fig:features} shows a complete list of the features that were used, ranked by descending importance.

\subsection{Training}
\label{subsect:training}
To train our classifier, we generated 1,100 different network topologies of varying complexity and heterogeneity so that balanced, labeled data relevant to both Tail Drop and AQM schemes were obtained. For each random network topology, we instantiated two consecutive, distinct Mininet simulations by varying the queue management technique on the bottleneck between Tail Drop and PIE. We simulated a long running TCP flow on each network from our known source to the sink and auxiliary flows from other randomly generated hosts to their respective sinks for 20 seconds. From the simulation, we gathered the resulting congestion window and RTT over time using \verb|tcpprobe| and \verb|ping| respectively. This data, as exemplified by the RTT plot in Fig.~\ref{fig:data}, were then used as labeled training data in our classifier.

%% file: evaluation.tex
\section{Evaluation}
\label{sec:evaluation}

Using the the data obtained from our simulations, we trained and benchmarked state-of-the-art classification algorithms such as Support Vector Machines (SVMs) \cite{svm}, Neural Networks (NNs) \cite{hornik1989multilayer}, Logistic Regression \cite{logistic}, and Random Forests, among others. We conducted k-fold cross validation (with $k = 10$) for each classifier on multiple data sets and found Neural Networks to be consistently outperforming all other tested classification techniques.

Having converged on using NNs as our classification algorithm, we employed a randomized search to optimize hyper-parameters by drawing from a previously specified parameter space distribution.
The parameter space included 1) NN architecture, i.e., the number of layers (1-4) and neurons (2-20) per layer,
2) $L_2$-Regularization parameter $\alpha$ penalizing large connection weights, and
3) solvers, including Stochastic Gradient Descent (SGD), ADAM, and Limited-memory Broyden Fletcher Goldfarb Shanno (LBFGS) algorithms.
The randomized search evaluates each parameter tuple with a score consisting of the average accuracy over 100 stratified cross-validations with $90\%$ training data and $10\%$ test data, to avoid biased parameter selection from overfitting. After 10,000 evaluations, the optimal hyper-parameters were found to be a NN topology with 1 hidden layer consisting of 14 neurons, a regularization of $\alpha = 0.5e-10$, and the LBFGS solver.

\subsection{Training Results}
\begin{figure}[t!]
\centering
\subfigure[Left Out Data]{%
\includegraphics[width=0.49\columnwidth]{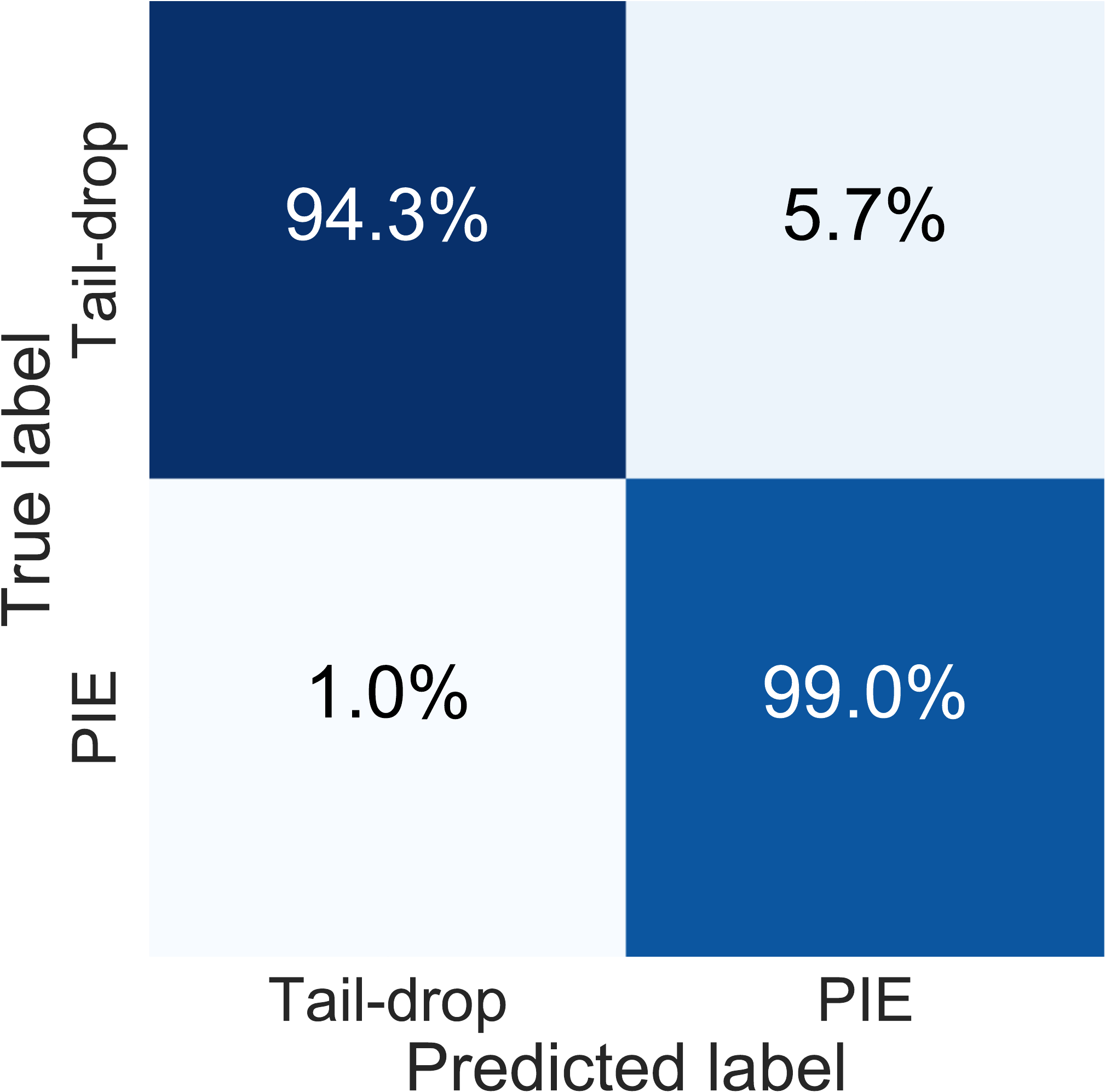}%
\label{fig:cm-left-out}}
%\vspace{3mm}
\subfigure[Complex Topologies]{%
\includegraphics[width=0.49\columnwidth]{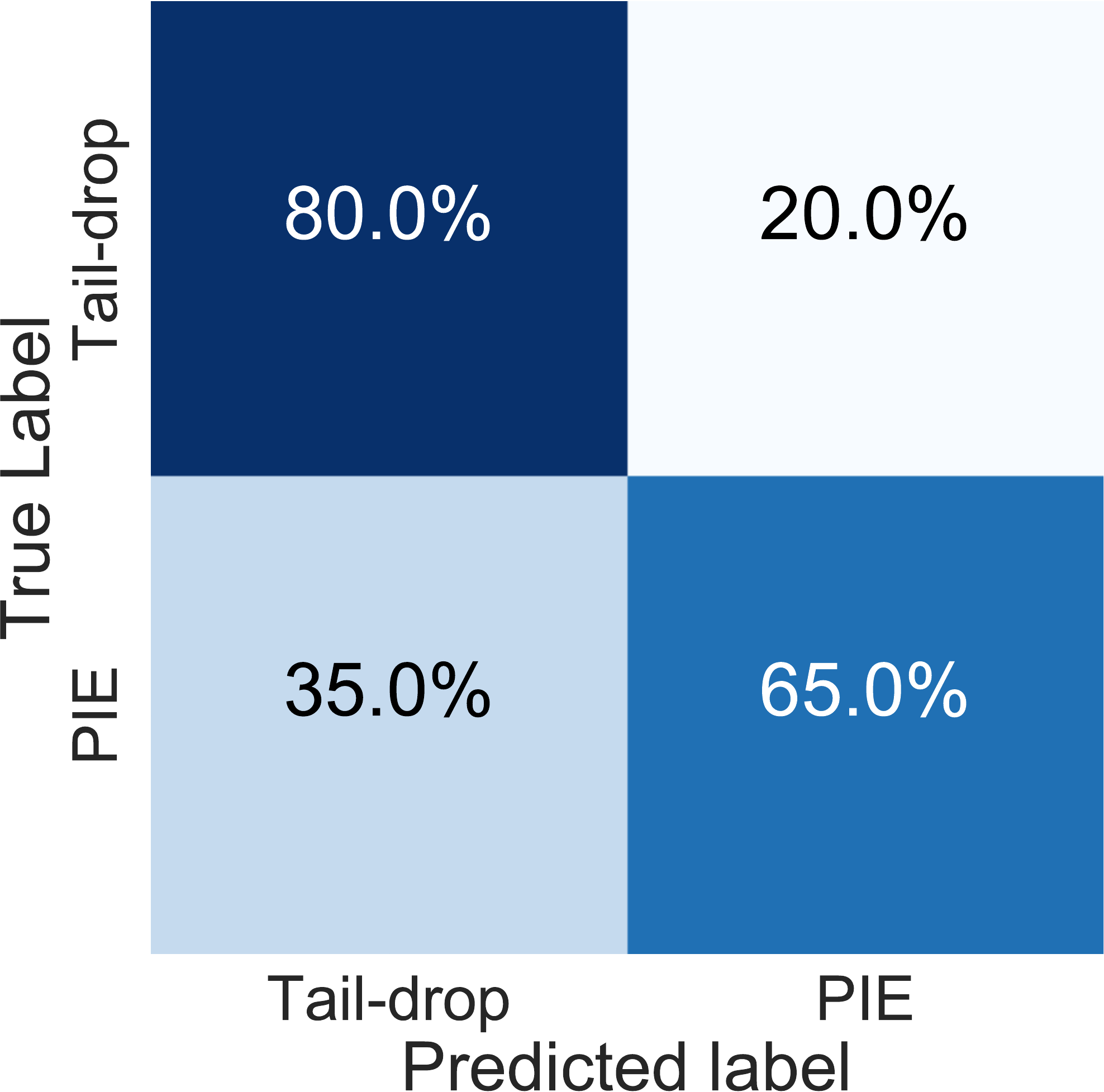}%
\label{fig:cm-complex}}
\caption{Confusion matrices for the two testing scenarios, (a) data left out of the training with the same topology complexity and (b) data generated using more complex topologies than used for training}
\label{fig:cm}
\end{figure}

From the 2,200 labeled data points that were generated, 200 were left out of the training to be used later for testing. After training the classifier and optimizing its hyper-parameters, we tested how well the classification generalized by evaluating its performance on the left-out data set. Fig.~\ref{fig:cm-left-out} shows a confusion matrix summarizing the results of this classification test. On a per-category basis, we achieve 94\% and 99\% prediction accuracy for correctly classifying the bottleneck router as Tail Drop or PIE respectively. Overall, we were able to accurately classify whether the bottleneck router was utilizing a Tail Drop or PIE as its queuing scheme 97\% of the time. This can be seen as an improvement over current state-of-the-art AQM detection algorithms, such as the recently proposed method by Bideh et al. \cite{tada}, which achieved an overall classification accuracy of 73\% on a serial bottleneck within a relatively simple network topology. In addition to achieving a higher classification accuracy than that of previous work, we also note that our algorithm was evaluated on significantly more complex and diverse network configurations, whereas the algorithm presented in \cite{tada} was evaluated against a single topology consisting of solely 3 hosts and 3 routers in total.

To thoroughly assess the dexterity of our algorithm in generating reliable predictions, we evaluated the classification accuracy of our algorithm on a data-set obtained from simulations of highly complex topologies with significantly more hosts, routers, and flows than the topologies that were used to train our algorithm. We emphasize the fact that our classifier was \emph{not} trained on any data from these more complicated scenarios.

Despite the challenges of this classification task, our classifier produced accurate labels for Tail Drop and PIE 80\% and 65\% of the time respectively, for an average classification accuracy of 73\% overall. The confusion matrix for these tests is shown in Fig.~\ref{fig:cm-complex}. This evaluation is evidence for our classifier's ability in generalizing well even to data obtained from highly complex network configurations that it was not explicitly trained to handle beforehand. Our favorable results demonstrate our algorithm's potential to be applied to real-world settings and achieve high prediction accuracy even against highly complex network configurations.

\subsection{Feature Importance}
\begin{figure}[t!]
\centering
\includegraphics[width=\columnwidth]{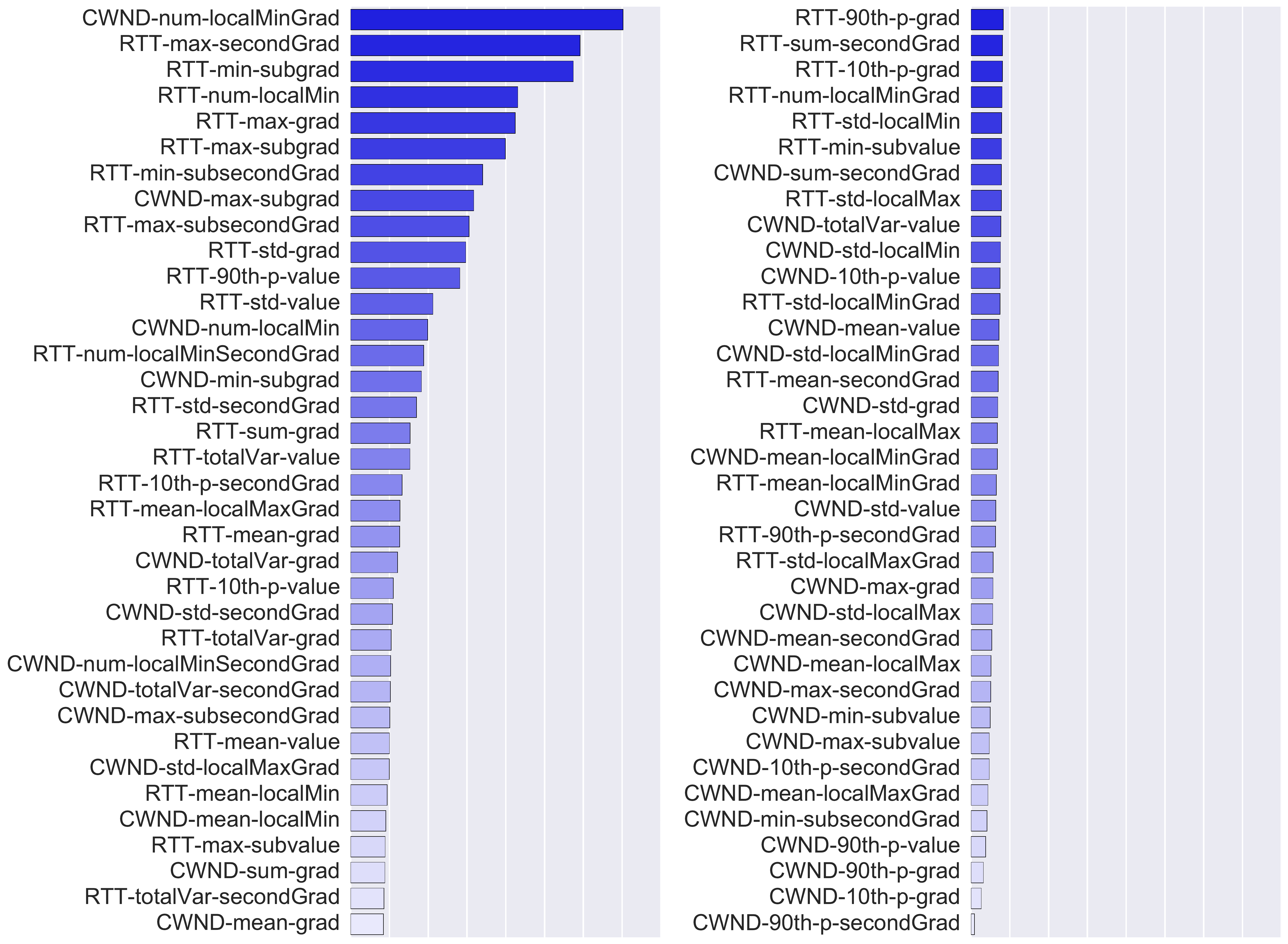}
\caption{Plot showing each feature's importance for classification. Note the high importance scores of features pertaining to RTT information and gradients' statistics.}
\label{fig:features}
\end{figure}

As introduced in Sec.~\ref{sec:approach}, our feature vector for a single data point was 72 dimensional, with 36 features each for RTT and CWND. The 72 distinguishing features for each data point mentioned in  Sec.~\ref{sec:approach} were motivated by our knowledge of the profound differences between Tail Drop and AQM schemes, as we covered in class \cite{class}. Fig.~\ref{fig:features} depicts a quantification of feature importance for each of the 72 features. By inspecting the highest ranking features, we obtain the insight that features pertaining to gradient-like statistics were the most distinguishing features and that data about the RTT over time were more helpful than the CWND data.

%% file: conclusion.tex
\section{Conclusion}
\label{sec:conclusion}
We presented a data-driven approach that leveraged a Neural Network for reliably determining whether a bottleneck router along a given path employs a Tail Drop or AQM scheme. We developed a framework for generating random network topologies and used Mininet to simulate a wide range of scenarios and obtain diverse training data. We identified distinguishing features by analyzing the intrinsic differences between AQM and Tail Drop and quantified the respective importance of each feature. Our feature importance analysis indicated that the most salient features were statistics pertaining to the gradient and RTT time-series data.

Our algorithm attained an accuracy of 97\% when evaluated against test data and generalized and performed well against data from highly complex topologies that it had not been exposed to before. We conjecture that our algorithm has potential to generalize to and provide reliable and accurate predictions in complex, real-world networks. Future work includes further training and evaluation on even more complex topologies and real-world data. We also plan to extend our method for classification of different AQM interfaces such as RED and PIE, in addition to the Tail Drop scheme.